\documentclass[vecphys]{svmult}

\usepackage{epsfig}
\usepackage{makeidx}         % allows index generation
\usepackage{graphicx}        % standard LaTeX graphics tool
\usepackage{multicol}        % used for the two-column index
\usepackage[bottom]{footmisc}% places footnotes at page bottom
\makeindex             % used for the subject index
\begin{document}
\title*{The beginning of string theory: a historical sketch}
\author{Paolo Di Vecchia\inst{1}\and
Adam Schwimmer\inst{2}}
\institute{Nordita, Blegdamsvej 17, 2100 Copenhagen {\O},
Denmark, \texttt{divecchi@nbi.dk} \and Weizmann
Institute, Rehovot 76100, Israel,
\texttt{adam.schwimmer@weizmann.ac.il}}
\maketitle

\begin{abstract}
In this note we follow the historical development of the ideas that led to
the formulation of String Theory. We start from
the inspired guess of Veneziano and its extension to the scattering of $N$
scalar particles, then we describe how the study
of its factorization properties allowed  to identify
the physical spectrum making the string worldsheet manifest 
and finally we discuss how the critical values of the
intercept of the Regge trajectory and of the critical dimension were
fixed to 1  and 26.
\end{abstract}

%Your text goes here. Separate text sections with the standard \LaTeX\
%sectioning commands.
\section{Introduction}
\label{intro}

The purpose of this note is to  follow the historical development of
the ideas that led to the formulation of  String Theory. As we will
discuss, the story consists of a remarkable succession of inspired
insights. First  Veneziano  guessed the form of the four point
function~\cite{VENE}. This was followed by its extension to
amplitudes with an arbitrary number of external legs. At this
point the dual resonance model was constructed. The subsequent
analysis of  its
factorization properties  allowed one to identify the full target
Hilbert space of physical states and its critical dimension by the
use of various consistency conditions. The natural interpretation of
the structure uncovered was that of a string propagating in
Minkovski space-time.
%in proper time, describing a  two dimensional worldsheet.

We want to stress that all this was achieved without the use of
a Lagrangian formulation but by implementing the basic principles of
S-matrix theory directly on the scattering amplitudes in a model
containing an infinite number of zero  width resonances. The new,
additional requirement was the Dolen-Horn-Schmid (DHS) 
duality~\cite{DHS} i.e. that  the sum of
resonances in one channel represents correctly the resonances in the
other channel.

As a result, the basic framework of Perturbative String Theory at
the operational level was well understood by 1971. Further progress
was achieved through the discovery of the Superstring and Space-time
Supersymmetry which led to tachyon free theories. Later some basic
concepts used before at a heuristic level like the origin of the
first class constraints necessary for making the spectrum unitary
and Lorentz invariant were put on a firm ground starting from the
action used in Ref.~\cite{POLY}.

Further conceptual developments like the connection between world
sheet conformal invariance and target space equations of motion
were only partially understood and had to wait for the first String Revolution
to get a more complete formulation. Finally
the relation between different String Theories through
dualities was the result  of the second String
Revolution.

In this note we will concentrate on the developments during the
period 1969-1972.

As we mentioned above three components entering the basic
structure of perturbative string theory i.e.:

\begin{itemize}
\item{the string world sheet}

\item{the physical spectrum and
vertex operators}

\item{the critical dimension}
\end{itemize}
were all correctly  identified by the end of 1972 and in this short note
we will limit ourselves to the description of the evolution of their
understanding. We will not cover  other very important
developments during the same period like e.g.  fermionic degrees
of freedom on the worldsheet (the Neveu-Schwarz-Ramond
formalism~\cite{NS,RAMOND}), compact degrees of freedom on the
world sheet leading to internal symmetries~\cite{HALP} and String
Field Theory in its light-cone formulation~\cite{SM1}.

We will follow the evolution of the ideas which led to the
understanding of  the three basic concepts above outlining the
most important conceptual jumps. Just the essential formulae will
be given referring
 for the detailed derivations to the accompanying paper~\cite{DIV}.
We will try to put in perspective the evolution of the ideas by
translating the guesses and insights in today's language and
understanding, as presented in the standard modern textbooks~\cite{POLCH}.
 We start with a brief reminder of the  developments on which
 the three breakthroughs mentioned above  were based.

\section{Prehistory: The discovery of the Dual Scattering Amplitudes}
 \label{Prehistory}

The first step which started the evolution of String Theory was
 the Veneziano Formula~\cite{VENE}.
By a historical accident Veneziano's formula refers to what
 is today Open String Theory. The analogous formula for
Closed String Theory guessed by Virasoro~\cite{VIRA}   was
generalized~\cite{SHAPI} and analysed later~\cite{DELGIU2} when
the basic structure of the open string was already understood. We
will follow the historical path and discuss only  Open String
Theory.

The formula guessed by Veneziano corresponds to  what we call
today the 2 to 2 scattering amplitude of the bosonic open string
tachyons:
\begin{eqnarray}
A (s, t, u) = A(s,t) + A(s, u) + A(t,u)
\label{astu}
\end{eqnarray}
where
\begin{eqnarray}
A(s,t) = \frac{\Gamma (- \alpha (s) )\Gamma (- \alpha (t) )}{
\Gamma (- \alpha (s) - \alpha (t) )} = \int_{0}^{1} dx x^{-
\alpha(s) -1} (1-x)^{-\alpha (t) -1} \label{ast}
\end{eqnarray}
and
\begin{eqnarray}
\label{rising} \alpha{(s)} = \alpha_{0} + \alpha ' s
\end{eqnarray}
is a linearly rising Regge trajectory.

The appearance of the free parameter $\alpha_{0}$ instead of the
 usual value $1$ will be discussed below. Moreover, in
 the Veneziano amplitude, as written above, there is no
 requirement that the external particles are the spin $0$
 particles on the leading trajectory $\alpha(s)$. Nevertheless we
 will continue to call the external particles "tachyons"  because they
have negative mass squared if we require them to be on the leading trajectory
 for $\alpha_0=1$.

In Veneziano's original approach  the amplitude was supposed to
 describe scattering of mesons due to strong interactions.
 The physical principles guiding Veneziano in his
guess were  the usual analyticity and crossing symmetry
requirements of the scattering amplitudes and a new principle,
the DHS duality~\cite{DHS}.

\begin{figure}[ht]
\epsfxsize = 6cm
\centerline{\epsfbox{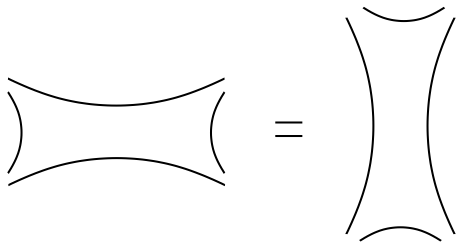}}
\caption{\small The duality diagram contains both s and t channel poles}
\label{dua}
\end{figure}

DHS duality was abstracted from a phenomenological study of
hadronic reactions and stated that the scattering amplitude could
be decomposed alternatively into a set of s-channel or t-channel
poles, each decomposition being complete, and containing, by
analytic continuation, the other. This was expressed by the
pictorial identity~\cite{HR}  in Fig. \ref{dua}.

In today's language it is qualitatively clear that the DHS
requirement is fulfilled if the amplitude is related to the
correlator of four vertex operators in a conformal field theory.
The two different decompositions which make explicit the pole
structure can be represented graphically by two ``duality
diagrams'' related by a continuous deformation and correspond to
the two possible decompositions in conformal blocks of the
conformal correlator. This happens if the conformal block is
translated into poles in Lorentz invariants constructed from the
space-time momenta. This basic feature of String Theory to which DHS
duality led, is very far from its phenomenological origin.
Ironically, it seems that present hadron scattering data~\cite{DL}
are not anymore in agreement with DHS duality which was
a feature related to the  energy range available at the time.

For the $N$-point function the DHS duality is generalized by
requiring that, for a fixed ordering of the external particles, the
amplitude can be represented by any one of the deformations of the
respective $N$-point duality diagram.
As described in Ref.~\cite{DIV}, one way to understand the
mechanism by which $A(s,t)$ satisfies the DHS duality is
to study its integral representation and identify
the two  mutually exclusive integration domains which produce
the poles  in the $s$ and in  $t$ channel respectively.
This is generalized
for the $N$-point function by writing it as a sum of terms, each one
corresponding to a given ordering of the external legs. Each term has
a $N-3$-dimensional integral representation. The  different deformations
 of the duality diagram are obtained from the singular contributions
  to the integral representation of mutually exclusive $N-3$
 dimensional integration regions.

Based on this idea the unique $N$-point function was
constructed in Ref.~\cite{NPARTI}:
\begin{eqnarray}
B_N = \prod_{i=2}^{N-2} \left[ \int_{0}^{1} d u_i u_{i}^{- \alpha
( s_i
  ) -1} (1 - u_i )^{\alpha_0 -1} \right] \prod_{i=2}^{N-2}
\prod_{j = i+1}^{N-1} (1
  - x_{ij} )^{2 \alpha' p_i \cdot p_j }
\label{bnb}
\end{eqnarray}
where
\begin{eqnarray}
s_i \equiv s_{1i}~~~;~~x_{ij} = u_i u_{i+1} \dots u_{j -1}
\label{con62}
\end{eqnarray}
\begin{eqnarray}
s_{ij} = - (p_i + p_{i+1} + \dots + p_j )^2 \label{sij}
\end{eqnarray}
and  $p_i, i=1,2,..,N,$ are the external momenta.  One requires
that the external scalar lies on the leading trajectory as explained
in Ref.~\cite{DIV}. Starting from this expression Koba and
Nielsen~\cite{KN} put it in the more symmetric $SL(2,R)$ invariant
form (see Ref.~\cite{DIV} for details)
\begin{eqnarray}
B_N = \int_{- \infty}^{\infty} d V (z) \prod_{(i ,j)} ( z_i ,
z_{i+1} , z_j , z_{j+1} )^{- \alpha ( s_{ij} )-1}
 \label{bnc}
\end{eqnarray}
where
\begin{eqnarray}
 d V (z) = \frac{ \prod_{1}^{N} \left[\theta (z_i - z_{i+1}) d z_i
   \right]}{\prod_{i=1}^{N} ( z_{i} - z_{i+2}
 ) d V_{abc}}~;~ d V_{abc} = \frac{dz_a dz_b dz_c}{(z_b -z_a
 )(z_c - z_b) (z_a - z_c)}
\label{dv}
\end{eqnarray}
and the variables $z_i$ are integrated along the real axis in a
cyclically ordered way: $z_1 \geq z_2 \dots \geq z_N$ with $a, b,
c$ arbitrarily chosen.

The $SL(2,R)$ group mentioned above acts on the integration
variables $z_i$ as a M{\"{o}}bius transformation:
\begin{eqnarray}
z_i \rightarrow \frac{\alpha  z_i + \beta}{\gamma  z_i +\delta
}~~;~~ i=1 \dots N~~;~~\alpha \delta - \beta \gamma =1
\label{proj}
\end{eqnarray}
Using the transformation in Eq. (\ref{proj}) for a fixed ordering, one
can relate amplitudes corresponding to circularly permuted
kinematical invariants and then adding terms for different
orderings one can show that all the requirements of crossing
symmetry are fulfilled. As we understand it today, the M{\"{o}}bius
transformations are related to globally defined reparametrizations
of the disk which leave invariant the metric up to a conformal
factor. This was the first  manifestation of the conformal
symmetry underlying the world sheet action of String Theory which
played an essential role in the understanding of the theory.

 The expression in Eq.  (\ref{bnc})
which was guessed as following from the principles mentioned above,
coincides (for $\alpha_0=1$) with the tree level scattering
amplitude of N open string tachyons obtained
 from calculating the open string path integral on a disk with
the insertion of N-tachyon vertex operators after mapping the disk
to the upper half plane.

The Koba-Nielsen form of the N-point function was the starting
point for the crucial developments which started in 1969.
There was a general feeling among the workers in the field that
the set of N-point functions  represent the result of a unique and consistent
underlying theory. While attempts to use the functions
to fit hadronic data continued, the search for this theory became
the major theoretical challenge.
One aspect which became immediately obvious was
the necessity to "unitarize" the theory: the presence of zero width poles
in the N-point functions showed that the amplitudes should be considered,
at best, as "tree diagrams" of an underlying, unknown theory
and "loop" diagrams should be added to them. A first
attempt~\cite{KSV} to write loop diagrams was by  using again
a generalized form of the DHS principle requiring a singularity
structure of the amplitudes consistent  with deformations of
duality diagrams involving loops. The existence of rather involved
integrals, found in  Ref~\cite{KSV},  which fulfill
the constraints, reinforced the belief in the existence of an underlying
theory. On the other hand, the ambiguities in the amplitudes
constructed originating in what we call today "the measure factors"
and the impossibility to verify the unitarity, reinforced the
necessity of understanding the basic underlying theory.

The approaches used  were conditioned by the development of the
 theoretical techniques at the time. Though the path integral
 formulation of Quantum Field Theory existed, it was  not well
 developed as a calculational tool. This was the case especially
 for gauge theories where the correct treatment of gauge
 symmetries  achieved a few years later by Faddeev-Popov
 did not exist. As a consequence Lagrangian methods based on an
 action were not very precise and involved some guess work at
 different stages. On the other hand, operatorial methods were well
 developed and through the Gupta-Bleuler treatment of
 QED as a prototype, even
 the  correct impositions of constraints corresponding to a
 gauge fixing (at least for the case when the ghosts are
 decoupled in today's language) were understood. We can
 roughly divide the  search for the
 underlying theory as the "Lagrangian approach" and the
"operatorial approach".

Since we will discuss later in more detail the operatorial approach
we start with a description of the evolution of the "Lagrangian" ideas.
Researchers following this path tried to guess the underlying
Lagrangian which
%following a certain interpretation
would lead to the N-point functions. This line was open by Nambu, Nielsen and
Susskind. Nambu~\cite{STRING1} and Susskind~\cite{SUSS} proposed
that the
 underlying dynamics of the dual N-point functions corresponds to
 a generalization of the Schwinger proper time formalism where
 a relativistic string is propagating in proper time. The
 equation of motion satisfied by the string coordinates  was the
 two-dimensional D'Alembert equation following from a
 linearized Lagrangian. Using plausible arguments they
 obtained  expressions similar to the N-point (tree) amplitudes.

Then Nielsen~\cite{NIELSEN} and immediately after
 Fairlie and Nielsen~\cite{FN}
used this linearized Lagrangian   for constructing the
"analogue model". The basic observation was that the momentum
dependence of the integrands in the Koba-Nielsen amplitudes and
their loop generalizations is related  to the energy of   two
dimensional electrostatic problems where the momenta are "charges"
located on the boundary. Then the electrostatic problem is solved
on a disk for the tree amplitude or on  an  higher genus two
dimensional surface described by the duality diagram corresponding
to the respective loop amplitude. We understand this result today
as a simple consequence of the fact that the $ik X(\sigma)$ factor
in the exponential of the vertex operator acts as a source for the
string coordinates whose propagator is the two dimensional Coulomb
kernel. Though the measure was not correctly reproduced, the
"analogue model" is important since it is the first appearance of
the two-dimensional world sheet in a mathematical role rather than
just as a picture in the duality diagram. This model is the
precursor of the path integral formulation of string theory that
was  understood completely only later.
%In particular the perturbative
%definition of the theory was proposed to be the sum over all the
%Riemann surfaces ~\cite{LOVE3,ALE,AA} .
Furthermore, the "analogue
model" motivated the generalization~\cite{SHAPI} of the Virasoro
amplitude~\cite{VIRA} and therefore the formulation of the Closed
String Theory by simply putting electrostatic sources on a sphere
instead on the boundary of a disk.

A non-linear action, proportional to the area  spanned  by the
string,
generalizing the non-linear one for the pointlike particle, was also
proposed by  Nambu and Goto in  Ref.s~\cite{NAMBU2,GOTO}.
But the consequences
of its non-linear structure, implying the invariance under an
arbitrary reparametrization of the world sheet coordinates, were only clarified
few years  later with  the treatment  of Ref.~\cite{GGRT}
that provides a
%more
rigorous derivation of the properties of the
generalized Veneziano model, though our present understanding of string
theory is mostly based on the action used in Ref.~\cite{POLY}.

The second approach that we will describe in  detail
in the next section, is based instead on the construction of an
operator formalism that made transparent the most important
properties of the model as the spectrum of physical states and their
scattering amplitudes and that historically has been  essential
for relating it to string theory in a completely satisfactory way.

%anchored
%the search for the underlying theory in the exact expression of
%the Koba-Nielsen amplitudes.

\section{The String World Sheet through Factorization of the
$N$-point amplitudes}
 \label{factorization}

The basic observation used in order to uncover the underlying theory
in the operatorial approach was that, having a set
of $N$-point functions satisfying the DHS
duality, crossing symmetry and tree level analyticity, does not define a
 consistent set of S-matrix elements unless
the different poles in the various channels can be shown
 to come from the same set of physical states, the residues being
factorized. This means that one should find a set of states  and a
set of three point couplings between these states such that any
expansion of a given ordering contribution to any of the $N$ point
functions is reproduced by the same set of states and couplings.

During 1969 there was an intensive activity in this program of
finding the universal set of states and couplings leading to
 factorization. We will describe in words the main steps in
 historical succession and then describe the complete solution
 as formulated in Ref.~\cite{FUVE2} at the end of 1969.
Through an explicit analysis of the residues of a given
pole in  Ref.s~\cite{FUVE1,BAMA}  it was shown that
 factorization can be achieved  by having an infinite number of
  intermediate states. An essential step was made in Ref.~\cite{FGV}
  where it was  proven
  that the spectrum is the Fock space of an infinite number of
  harmonic oscillators. The authors of  Ref.~\cite{FGV} gave general
   formulae for the masses of
  the states in terms of occupation numbers and for the couplings
  of the external tachyons to arbitrary pairs of states in terms
  of matrix elements of vertex operators depending on the harmonic
  oscillator degrees of freedom.  An important result of
Ref.~\cite{FGV} was the discovery of the existence
of the Hagedorn temperature in the
theory, a basic feature characterizing String Theories.

We describe now the solution of the factorization problem
following  Ref.~\cite{FUVE2}. One starts defining
the  operator $Q_\mu(z)$ by:
\begin{eqnarray}
Q_{\mu} (z) = Q^{(+)} (z)  +  Q^{(0)} (z) + Q^{(-)} (z)
\label{qfv}
\end{eqnarray}
where
\[Q^{(+)} =  i \sqrt{2 \alpha'}  \sum_{n=1}^{\infty}
\frac{a_n}{\sqrt{n}} z^{-n}~~;~~ Q^{(-)} =  -i \sqrt{2 \alpha'}
\sum_{n=1}^{\infty} \frac{a^{\dagger}_{n}}{\sqrt{n}}
 z^{ n}
 \]
 \begin{eqnarray}
 Q^{(0)} = {\hat{q}} - 2 i  \alpha' {\hat{p}} \log z
 \label{qfva}
\end{eqnarray}
and the vertex operator by
\begin{eqnarray}
{{V}} ( z ; p ) = : {e}^{i p \cdot Q (z)} : \equiv {e}^{i p \cdot
Q^{(-)} (z) } {e}^{i p {\hat{q}}}{e}^{ + 2 \alpha' {\hat{p}} \log z}
{e}^{i p \cdot Q^{(+) (z) }}
\label{vertope}
\end{eqnarray}
Then it was shown~\cite{FUVE2}  that
the integrand  of the Koba-Nielsen $N$-point function is related
to  the Fock space vacuum matrix element of the product of vertex
operators:
\begin{eqnarray}
\langle 0, 0 | \prod_{i=1}^{N} {{V}} ( z_i , p_i)   |0, 0\rangle =
 \prod_{i >j} ( z_i - z_j )^{2 \alpha' p_i \cdot p_j} (2 \pi)^4
 \delta^{(4)} ( \sum_{i=1}^{N} p_i )
\label{ANc}
\end{eqnarray}
In order to obtain exactly the Koba-Nielsen expression one has to
deal carefully with the fixing of three of the $z$ variables. This
is done by   extracting the $z$ dependence of the vertex operators
using the identity:
 \begin{eqnarray}
z^{L_0} V (1  , p) z^{-L_0} = V ( z, p) z^{\alpha_0}
 \label{dila4}
\end{eqnarray}
 where $L_0$ is the operator:
\begin{eqnarray}
L_0 = \alpha' {\hat{p}}^2 + \sum_{n=1}^{\infty} n a_{n}^{\dagger} \cdot
a_n
\label{elzero}
\end{eqnarray}
Choosing three consecutive values of $z_i$ to be fixed:
\begin{eqnarray} z_a = z_1 = \infty~~;~~ z_b = z_2 =
1~~;~~ z_c = z_N =0
\label{cho8}
\end{eqnarray}
the Koba-Nielsen amplitude can be rewritten in the operator
language as:
\begin{eqnarray}
A_{N} \equiv \langle 0, p_1 | V (1, p_2) D V ( 1, p_3 ) \dots  D V
(1, p_{N-1} ) | 0, p_N \rangle
\label{bng}
\end{eqnarray}
where the "propagator" $D$ is equal to:
\begin{eqnarray}
D = \int_{0}^{1} d x x^{L_0 -1 - \alpha_0 } (1 -x )^{\alpha_0 -1}
= \frac{  \Gamma ( L_0 - \alpha_{0} ) \Gamma ( \alpha_{0}  )}{
\Gamma ( L_0  ) } \label{propa5}
\end{eqnarray}
and the states (using what we understand today  as "operator-state
correspondence") are defined as:
\begin{eqnarray}
 \lim_{z \rightarrow 0} V
(z;p) | 0,0\rangle \equiv | 0 ; p \rangle ~~;~~ \langle 0; 0 |
\lim_{z \rightarrow \infty}  z^{2 \alpha_0} V (z; p) = \langle 0,
p | \label{limi67}
\end{eqnarray}
The $z_i$ integrations of the Koba-Nielsen formula which were absorbed in the
definition  in Eq. (\ref{propa5}) are translated into
integrations over the "proper times" $x_i$ appearing in the propagators.

This provides an explicit solution to the factorization. In fact, one can
insert  between each $V$ and $ D$ a complete set of states of the space
spanned  by  the harmonic
oscillators (Fock space) appearing in $Q(z)$.
Since $D$ is diagonal in the basis of occupation numbers, poles will
appear at $\alpha(s)=0,1,2,...$   with factorized residues related
in a universal fashion to the matrix elements of the vertex
operators.

This solution to the factorization
 problem was the crucial step in the development of String Theory
 since, from
now on, the  $N$-point functions were clearly related to a
theory in which the set of space-time fields  is labeled by the
 states in the Fock space on which the $Q_\mu$ fields are
realized. The $Q_\mu$ fields are, of course, the open string
coordinate fields $X^{\mu} (\sigma, \tau)$ in $d$ space-time
dimensions for $\mu=0,1,2,..,d-1$, computed at
the endpoint of the string coordinate $\sigma =0$, where $z$ is
related to the other string coordinate $\tau$ by $ z=e^{i \tau}$. They
are Heisenberg operators, their dependence on the
world-sheet coordinates  $\sigma$ and $\tau$ follows from the fact
that they are  solutions of an equation of motion following from
a free linearized Lagrangian. However, as it is described above the
Lagrangian was not used in the derivation, the various expressions
being obtained by a rewriting of the N-point amplitudes. While the
linear spacing between the poles of the Veneziano formula was
suggestive of some underlying harmonic oscillator type structure
 only the solution of the
factorization problem unveiled the true structure of the theory,
i.e. an infinite number of oscillators assembled into a set
 of fields $Q_\mu$ living on a two-dimensional world sheet.

The vertex operators for the emission of tachyons represent
insertions on the boundary (for open string theories) of the
two-dimensional world sheet.
Of course the relation in Eq. (\ref{bng}) is the way in which
scattering amplitudes are obtained in String Theory starting from
the matrix element of products of vertex operators.
%in the operator formalism
%representing the path integral over the world sheet
The historical way was exactly the opposite i.e.
given the Koba-Nielsen formula,   the operators whose matrix
elements reproduce the formula were correctly guessed identifying
the Hilbert space.
%and through it the path integral.
Now the fulfillment of the DHS requirements became  natural: the $Q_\mu$
are massless two dimensional fields defining a two-dimensional
conformal theory and the $N$-point functions are related to
integrals of correlators of the vertex operators in the $SL(2,R)$
invariant vacuum. The integration over the $z$ variables required
by the Koba-Nielsen formula, in order to produce the poles in
$\alpha(s_{ij})$, is related to the integration over the "proper
times" after the mapping of the disk into the upper half  plane. The fact that
this particular expression is special to a particular gauge (at
that time called the orthonormal gauge) was already understood
during the first period of String Theory, but  it became more
transparent and rigorous after  Polyakov's seminal paper~\cite{POLY}.

Having the decomposition of the amplitudes in "vertices" and
"propagators" allows the calculation of loop diagrams by gluing
them and taking traces for the loops. The loop diagrams are
necessary for producing an S-matrix consistent with unitarity.
In this way, one obtained already in 1970 the correct
expression in the Schottky parametrization of  quantities defined on
a Riemann surface as the period matrix, the abelian differentials and
the Green's functions~\cite{LOVE1b,ALE,AA}. However,
the correct measure of integration in the multiloops was not known at the
time since it requires the understanding of ghost contributions.
It is clear now that these operatorial expressions in the covariant
gauge are the same as those obtained by performing the path integral
of the string Lagrangian over the appropriate world sheet.
%in the modern formulation.

We know today that the restrictions on the operators $V$ and $D$
which can be used
follow from a correct gauge fixing of the string Lagrangian. In the
absence of a Lagrangian again the correct restrictions on $V$
and $D$ were found by a rather tortuous path (from today point of
view) which we are going now to describe.

The expressions used above differ from the ones used in the modern
formulation in two respects:

i)The vertex
operators used  were defined for a conformal weight $\alpha' k^2$. This
value, related to the mass squared of the open string tachyon, is given in
terms of the arbitrary parameter $\alpha_0$: $\alpha_0 = \alpha' k^2$.
% by $-\alpha_0 \over \alpha^{'}$.

ii)The dimension $d$ of space-time i.e. the number of
string coordinates was left free.

\section{The Virasoro Conditions}
\label{Virasoro}
We start this section by reminding  the reader
how the two points mentioned at the end of the previous section
are understood today. The  starting point  today for the bosonic
string theory is   the $\sigma$-model
action (the action used in Ref.~\cite{POLY})
that, at the classical level, couples the string coordinates to
the two dimensional
world sheet metric in a diffeomorphism  and Weyl invariant manner.
Then the requirement that these two ``gauge symmetries'' (diffeomorphism and
Weyl) are not anomalous in the quantum theory fixes the space-time
dimension to the value $d=26$ for the bosonic string.

Once  the two ``gauge symmetries''  are respected at the quantum
level,  the standard Faddeev-Popov procedure can be applied,
in principle in  an arbitrary gauge, and
 a consistent quantization can be performed giving the physical
 states/operators  in the gauge chosen. The states/operators in different
 gauges are isomorphic leading to the same results when gauge invariant
 correlators are calculated. In particular, by
 choosing a covariant gauge the Lorentz invariance of the theory
 follows automatically, while the unitarity of the theory is not obvious.
 On the other hand,  by choosing an explicitly unitary gauge (the
 light-cone gauge) the unitarity of the theory is completely manifest, while
 the Lorentz invariance has to be checked.
 In the covariant gauge the physical
 states correspond to operators with dimension $1$ for the open
 string and $(1,1)$ for the closed string. This fixes the leading
 Regge trajectories to have intercept $\alpha_0=1$ or $\alpha_0=2$
 for the open and closed strings, respectively.
In a ``physical'' gauge, as the light-cone gauge,
the states which are now "transverse"  correspond to
cohomologically equivalent families in the covariant gauge.

%We have described above the present procedure for quantizing
%the bosonic string. However it must also be said that, in practice, one
%can  invert the logic outlined above and fix the
% normal ordering constants
%Regge intercept and the space-time dimension in the light-cone gauge
%by requiring that the Lorentz algebra be obeyed at the quantum
%level. This is, in fact, 
%the way followed in the early days of string theory when the procedure 
% described above was not yet known and this,
% of course, has led to the above values of the critical dimension and
%intercepts. Actually, to be more precise, the point of view
%expressed above has been  essential, when we quantize
%the bosonic string in a covariant gauge, only  in order to
%compute the correct integration measure for multiloop amplitudes. It
%has not played, in practice, any  significant role in the light-cone 
%gauge where the
%Regge slope and the space-time dimension have been correctly
%determined by
%imposing the closure of the Lorentz algebra~\cite{GGRT}.
Actually the BRST approach outlined above is essential for the
bosonic string only in order to compute, with manifestly Lorentz
covariant methods, the correct integration measure for multiloop
amplitudes. The understanding reached at the end of 1972 was almost
complete including tree and loop diagrams.

We want to stress here once more that none of the ideas based on the
BRST invariant approach (including the  $\sigma$-model action) were
known in the early days of string theory. The Nambu-Goto action
was known, but it was not really understood how to use it rigorously
for deriving all the properties obtained using the operator
formalism. One had to use alternative methods
which amazingly enough led to the correct results. This is what we
are going to explain below.

Before we proceed, let us notice that, from the present point of
view, the description done in the previous section involved just a
conformal theory of $d$ massless fields. Of course in such a theory
any vertex operator is legal and the correlators of vertex operators
on the $SL(2,R)$ invariant vacuum have the block decomposition
properties even after integrating over their "proper time"
coordinates. Interestingly, even without the understanding that a
consistent String Theory
 should be the gauge fixed version of a Weyl anomaly free theory
the way to make the theory consistent by restricting i) and ii)
was correctly guessed. This was done by looking for some "gauge"
conditions that could help in decoupling the negative norm states, required
by manifest Lorentz covariance, from the spectrum of the physical
states pretty much in analogy with what was known to happen in QED.
We start discussing the way in which the correct gauge conditions were
discovered.

In Ref.~\cite{FUVE1} it was pointed out that the residues of the
poles on which the amplitude is factorized are not positive definite
simply due to the presence of the time components of the oscillators
which in the operator formulation lead to a negative contribution to
the scalar  product. As a possible way out from this inconsistency
of the theory,  linear relations between the residues were uncovered
leading to the decoupling of some Fock space states  from the
amplitude. The basic driving idea was  that the situation here was
analogous  to the Gupta-Bleuler quantization of QED. As in QED
where the Lorentz condition was imposed to characterize the subspace
of the physical states,   here also
 some "gauge" conditions, that later on were understood to
be due to some  first class constraints, were imposed on the spectrum
which would eliminate the negative norm states.

 In this way
one managed to get the correct result without having to fix the
gauge of the diffeomorphisms and Weyl invariance and to introduce
 the $b,c$ ghost system.  This has been possible because the ghosts
are decoupled from the string coordinates. As a
consequence, the nontrivial BRST cohomology can be realized in
terms of the string coordinates only, the ghost ground state not
being excited and, for tree diagrams at least, one can calculate
consistently using the string coordinates restricted by the first
class constraints.

    The correct final
answer was reached following  a rather tortuous, but
physical and at that time intuitive path.

We start  describing   the
 linear relations~\cite{FUVE3}  mentioned above.
  In the operatorial formalism there is a realization~\cite{GLIOZZI,CMR}
of the M{\"{o}}bius transformations in Eq. (\ref{proj}) in terms of the infinite set
of harmonic oscillators. This  SL(2,R) algebra
has a simple action on the vertex operators and  annihilates the
vacuum. Its generators  $L_1,L_0,L_{-1}$ are:
\begin{eqnarray}
L_0 = \alpha' {\hat{p}}^2 + \sum_{n=1}^{\infty} n a_{n}^{\dagger} \cdot
a_n~;~ L_1 = \sqrt{2 \alpha'} {\hat{p}} \cdot a_1 +
\sum_{n=1}^{\infty} \sqrt{n(n+1)} a_{n+1} \cdot a_{n}^{\dagger}
\label{elle}
\end{eqnarray}
and
\begin{eqnarray}
L_{-1} = L_{1}^{\dagger} = \sqrt{2 \alpha'} {\hat{p}} \cdot
a_{1}^{\dagger} + \sum_{n=1}^{\infty} \sqrt{n(n+1)}
a_{n+1}^{\dagger}  \cdot a_{n} \label{L-1}
\end{eqnarray}
We recognize, of course, the central extension free $SL(2,R)$
subalgebra of the Virasoro algebra which acts as a symmetry on an
arbitrary CFT correlator, provided it is evaluated on the
$SL(2,R)$ invariant vacuum. We remind the reader, however, that
the algebra of the Virasoro operators and more generally two
dimensional conformal
field theories were not known at the time. Their understanding was a
result of the developments we are describing. The $SL(2,R)$
subalgebra generates the M{\"{o}}bius group of the finite
transformations of $z$:
\begin{eqnarray}
z'=\frac {\alpha z+\beta} {\gamma z+\delta}
 \label{mobius}
\end{eqnarray}
where $\alpha \delta -\beta\gamma =1$.
 The vertex operators have the standard transformation properties
 under the M{\"{o}}bius group corresponding to the weight $L_0=\alpha' p^2$.
 In the expectation value in Eq. (\ref{bng}) the information that $z_a$
 is fixed appears only through  the "bra" vector on the l.h.s. of the
 matrix element. Therefore the r.h.s. has  a residual symmetry,
 the subgroup of the M{\"{o}}bius group which leaves the fixed
 $z_b=1,z_c=0$ unchanged :
 \begin{eqnarray}
z' = \frac{z}{1 - \alpha (z-1)} = z + \alpha ( z^2 - z ) + o(\alpha^2 )
\label{prota}
\end{eqnarray}
This subgroup is generated by
\begin{eqnarray}
 W_1 = L_1 - L_0
  \label{gene}
\end{eqnarray}
Since the "ket" on the r.h.s. is left invariant by the subgroup in Eq.
(\ref{prota}) we obtain :
\begin{eqnarray}
W_1 | p_{(1, M)}   \rangle =0
 \label{genee}
\end{eqnarray}
 where :
 \begin{eqnarray}
| p_{1, M) }   \rangle = V (1, p_{M}) D \dots V ( 1, p_{2} ) |
p_{1} ,0\rangle
 \label{factb}
\end{eqnarray}
independently on the number of $V D $ insertions.
 Clearly, one gauge condition $W_1$ is not enough to project out
 all the negative norm states and additional conditions were
 searched for. We remark that Eq. (\ref{genee}) is not a consequence of
 any gauge symmetry, being valid in any CFT for  vertex operators
 of arbitrary  dimensions provided the vertex operators are
 inserted at the value $z=1$. Nevertheless  following the pattern
 that led to   Eq. (\ref{gene}),
 Virasoro~\cite{VIRA2} realized that, if $\alpha_0=1$, the state in
 Eq. (\ref{factb}) is annihilated by
an infinite set of "gauge" operators:
\begin{eqnarray}
W_n  | p_{1, M) }   \rangle =0~~;~~n=1,2,3, \dots
\label{wn1}
\end{eqnarray}
where
\begin{eqnarray}
W_n = L_n - L_0  - (n-1)
\label{wn}
\end{eqnarray}
with :
\[
L_n = \sqrt{2 \alpha' n} {\hat{p}} \cdot a_{n} +
\sum_{m=1}^{\infty} \sqrt{m (n+m)} a_{n+m} \cdot a_{m}  +
\]
\begin{eqnarray}
+ \frac{1}{2} \sum_{m=1}^{n} \sqrt{m ( n-m)} a_{m-n} \cdot a_m~~~;
n \geq 0~~~L_{-n} = L_{n}^{\dagger}
\label{ellen}
\end{eqnarray}
The "gauge" conditions in Eq. (\ref{wn1}) imply the following
equations for the on shell physical states of the
generalized Veneziano model~\cite{DELGIU}
\begin{eqnarray}
( L_0 -1) | Phys \rangle =  L_n | Phys \rangle =0 ~~~;~~~  n=1,2, \dots
 \label{wnn}
\end{eqnarray}
These are exactly the constraints  following from the diffeomorphism
and Weyl symmetry of the action in presence of a two dimensional
metric after the gauge fixing which eliminates completely the
metric. These constraints annihilate the intermediate states in Eq.
(\ref{bng}), that are not physical,  as we know from the now
standard gauge fixing-BRST procedure~\cite{POLCH}. We postpone the
discussion of the exact conditions under which the constraints
eliminate the negative norm states to the next section since it is
closely tied to the recognition of the critical dimension. In
conclusion, the correct results were obtained at the tree level
without needing to know the underlying Lagrangian and to introduce
the ghost degrees of freedom. What is more amazing is that also
the
 one-loop measure was correctly obtained by using the Brink-Olive
operator, that projected in the subspace of physical
states~\cite{BO}. The correct measure for the multiloop
amplitudes was  determined much later, although it would have been
possible, in principle, to determine it by extending the procedure
of Brink and Olive to multiloops.

Once the intercept $\alpha_0$ got fixed
to $1$ it became clear that the first state on the leading
trajectory is a tachyon; its consistent removal was achieved  only
with the discovery of the superstring and the GSO
projection~\cite{GSO}. Imposing the infinite set of Virasoro
constraints on the vertex operators corresponds in today's
language, that was already used in Ref.~\cite{CAMPAGNA},
to the requirement that the vertex operators should be  primary
fields with dimension $1$~\cite{CAMPAGNA}.
Projecting from the Fock space the
states which are annihilated by all the Virasoro constraints and
eliminating the zero norm states following the
procedure explained in  Ref.~\cite{DELGIU},
defines the physical Hilbert space which should have positive
norm.

 Shortly after Virasoro found the constraints (\ref{wn}) it was realized that
 the $L_n$ operators  are the generators of the conformal group
 in $d=2$~\cite{FUVE3}. The
 full algebra of the group including the central extension present
 in the commutator of $L_n$ with $L_{-n}$ was correctly worked out
only somehow later~\cite{WEIS}~\footnote{See note added in proof of
Ref.~\cite{FUVE3}.}. In this way the algebra of the Virasoro operators was
 established and became the basic algebraic structure underlying
  two-dimensional CFT and String Theory.
The  central extension discovered  by Weis~\cite{WEIS}
which is understood today as a manifestation of the conformal
anomaly~\cite{POLY}, has far reaching consequences which we are
going to discuss now.

 \section{The Critical Dimension}
 \label{Critical}

\begin{figure}
\begin{center}
\includegraphics{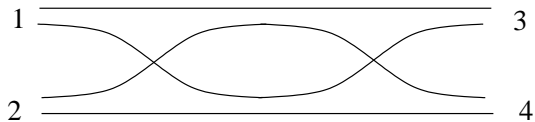}\caption{The doubly twisted open string diagram.}
\label{FRdiagram}
\end{center}
\end{figure}

 The discovery of the critical dimension with its various
 manifestations shows the serendipity characteristic of this first
 period of String Theory.
 Since, as we know it today, the existence of the critical
 dimension is a consequence  of the conformal
 anomaly cancellation
 between the string coordinates fields and the $b,c$ ghost system,
%as we  have discussed at the beginning of the previous section,
 it is clear that in   the absence
 of the understanding of the coupling to two dimensional metrics
 and its gauge fixing which leads to the ghosts, the critical dimension could
 manifest itself only  through its "side effects"
i.e. various consistency conditions of the theory.
 The first calculation pointing to the existence of the critical
 dimension was done by Lovelace~\cite{LOVE1}. He calculated
the non-planar loop with a number of tachyons as external particles,
represented in Fig.\ref{FRdiagram}.

\begin{figure}
\begin{center}
\includegraphics{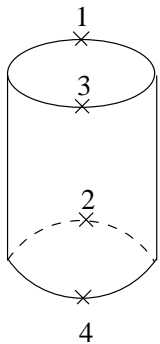}
\caption{
%The closed string coupled to open strings diagram
The diagram of Fig. \ref{FRdiagram} in the closed string channel.}
\label{closed}\end{center}
\end{figure}

 This diagram was proposed
 earlier~\cite{FR} as a model for the "Pomeron" which dominates the
 high energy elastic scattering amplitude of hadrons and therefore according
 to the lore of the time  was described
  as the Regge pole in the t-channel
 with the highest intercept. 
The diagram was first calculated in  d=4 ~\cite{PRELOVE} and
 the analytic structure showed the presence of a branch 
cut in the t-channel. In Lovelace's  calculation the dimension of
 space-time $d$ and the effective number of dimensions going
 around in the loop $d'$, were left as free parameters. It was
 understood at the time that only the physical degrees of freedom
 which obey the Virasoro gauge conditions circulate in the loops
 but the exact way to implement this fact was not understood~\footnote{This
 was clarified few years later by Brink and Olive~\cite{BO}
inserting in the loop
 the operator that projected into the space of physical states.}. The
 result of the calculation showed that  the singularity in the
 $t$-channel became a pole only when $d=26$ and $d'=24$ and in
 this case the intercept of the "Pomeron" Regge trajectory is $2$.
  We understand this result today as a consequence of the
  conformal invariance of the theory: by a continuous deformation
  of the world sheet the diagram in Fig,\ref{FRdiagram} can be brought
to the form in Fig.\ref{closed}.

 Now it is clear that one has a tree diagram,
   in the $t$-channel a closed string (the cylinder) being
   exchanged with the open string tachyons being coupled to the
   upper and lower disks. As a consequence 
in the t-channel one should have only poles. 
However, the conformal deformation of
   the world sheet on which the above expectation is based is
   valid only when conformal transformations act as expected
   classically i.e. no anomaly is present implying $d=26$.
    In addition, we know
   today that the $b,c$ ghosts circulating in the loop cancel the
   contribution of two of the space-time string coordinates
   leading to $d'=24$. Finally the intercept $2$ is the one
   required by the correct gauge fixing for the closed
   string. We identify nowadays the trajectory in the $t$-channel
   with the graviton and not the Pomeron though the connection may
   come back to haunt us~\cite{CHUNG}. In the critical case the
   couplings of the open strings can be factorized and a
   consistent open-closed theory can be constructed~\cite{POME,MIX}.

  Further  evidence for the existence of the critical dimension
  came from a close examination of the physical spectrum i.e. the
  Hilbert space left after the infinite set of Virasoro conditions
  are imposed on the Fock space. In Ref.s~\cite{GT,BROWER}
  it was shown that  the physical spectrum i.e. the ensemble of
Fock space states which satisfy the conditions in Eq. (\ref{wnn}) has a
positive definite scalar product (it is "ghost free") only
  when $d\leq 26$. Of course, if the spectrum is ghost free for
  $d=26$, it is a fortiori so also for $d < 26$.  In order to prove
  the "no ghost theorem" for $d=26$  the manipulations used in
  Ref.~\cite{GT} are very similar to the modern ones based on the
  BRST formalism and which are valid provided that the BRST operator
  $Q$ obeys at the quantum level $Q^2=0$ 
which requires $d=26$. As a corollary of their
  proof Goddard and Thorn showed that the DDF~\cite {DDF}
  states form a basis for the physical Hilbert space.

  This leads to a third manifestation~\cite{GGRT}
  of the critical dimension which is already very close to our
  modern understanding. Though the starting point in Ref.~\cite{GGRT}
  is the Nambu-Goto action the final results correspond to a
  correct quantization in light-cone~\cite{POLCH} and in
  covariant gauge~\cite{POLCH} of the $\sigma$-model action.
  The DDF states are isomorphic to the states
  in the light-cone gauge which live in a Hilbert space which has
  an explicitely positive definite scalar product. The light-cone
  gauge is, therefore,  unitary, however Lorentz
  invariance is not explicit. On the other hand, in the covariant
  gauge Lorentz invariance is explicit but unitarity is valid only on the
  physical Hilbert space after the imposition of the conditions in
  Eq. (\ref{wnn}). In our modern understanding the two gauges being
  equivalent at the critical dimension insures without further
  proof that the spectrum is both unitary and Lorentz invariant.
  However, at the time one had to
  prove  explicitly that on the spectrum in the light-cone
  gauge the Lorentz algebra is fully realized. By constructing all
  the Lorentz generators in  Ref.~\cite{GGRT}  it was shown that the algebra
  closes correctly  only if $d=26$. The treatment in Ref.~\cite{GGRT}
  is already very close to the modern one and completely satisfactory
  for computing string diagrams in the light-cone gauge. The modern
  BRST treatment from a calculational point of view just fixes the
  measure for the multiloop amplitudes.

  We mention finally an interesting interpretation of the central extension
  (and implicitly of the critical dimension) given by Brink and Nielsen in
  Ref.~\cite{BNIELSEN}.   They related the central extension to the Casimir
  energy of the
  string. In our understanding of today this is simply the fact
  that, transforming $L_0$ to the strip (or cylinder for the closed
  string) coordinates, an additional term proportional to the
  central extension appears. This argument was later generalized to an
  arbitrary CFT in Ref.~\cite{CARDY} giving a relation between the
  central extension and energies on finite geometries.

  \section{Conclusions}
   \label{Conclusions}

    In this history-oriented note we briefly reviewed some of the
    developments that led to what we call today "String Theory".
    At the end of 1972 a complete theory existed (as summarized in
    Ref.~\cite{GGRT}) which, except for the existence of the tachyon, was
    consistent. Its perturbative spectrum and the precise rules
    for calculating perturbatively scattering amplitudes
    were completely understood in the operator formalism. The
    theory is unitary and Lorentz invariant for $\alpha_0=1$
    and $d=26$. All this was obtained starting from a rather
    strange physical motivation and involved a long chain of
    beautiful conceptual insights and guesses. The impressive
    theoretical structure created in the years 1969-1972
     and further intensively developed
    during the last twenty five years  continues to be
    at the forefront of Theoretical Physics. We dedicate this
    contribution to Gabriele Veneziano who played a leading role
    in the developments we described.

%\input{referencrete}

 %%%%%%%%%%%%%%%%%%%%%%%% referenc.tex %%%%%%%%%%%%%%%%%%%%%%%%%%%%%%
% sample references
% "physics"
%
% Use this file as a template for your own input.
%
%%%%%%%%%%%%%%%%%%%%%%%% Springer-Verlag %%%%%%%%%%%%%%%%%%%%%%%%%%

%
% BibTeX users please use
% \bibliographystyle{}
% \bibliography{}

\end{document}